\documentclass[twocolumn,showpacs,preprintnumbers,amsmath,amssymb]{revtex4}

\usepackage{graphicx}
\usepackage{amssymb}
\usepackage{dcolumn}

\begin{document}

\draft
\title{Flux fluctuations in a multi-random-walker model and surface growth dynamics}
\author{S. Y. Yoon, Byoung-sun Ahn}
\author{Yup Kim} \email{ykim@khu.ac.kr  }
\affiliation{Department of Physics and
Research Institute for Basic Sciences, Kyung Hee University, Seoul
130-701, Korea}


\begin{abstract}

We study the dynamics of visitation flux in a multi-random-walker
model by comparison to surface growth dynamics in which one random
walker drops a particle to a node at each time the walker visits
the node. In each independent experiment (trial or day) for the
multi-random-walker model, the number of walkers are randomly
chosen from the uniform distribution $[\langle N_{RW} \rangle
-\triangle N_{RW} , \langle N_{RW} \rangle +\triangle N_{RW} ]$.
The averaged fluctuation $\overline {\sigma} ({T_{RW}})$ of the
visitations over all nodes $i$ and independent experiments is
shown to satisfy the power-law dependence on the walk step
$T_{RW}$ as $\overline {\sigma} ({T_{RW}})\simeq {T_{RW}}^\beta$.
Furthermore two distinct values of the exponent $\beta$ are found
on a scale-free network, a random network and regular lattices.
One is $\beta_i$, which is equal to the growth exponent $\beta$
for the surface fluctuation $W$ in one-random-walker model, and
the other is $\beta=1$. $\beta_i$ is found for small $\triangle
N_{RW}$ or for the system governed by the internal intrinsic
dynamics. In contrast $\beta=1$ is found for large $\triangle
N_{RW}$ or for the system governed by the external flux
variations. The implications of our results to the recent studies
on fluctuation dynamics of the nodes on networks are discussed.

\end{abstract}

 \maketitle


Recent advances of network theories \cite{AB} have elucidated that
the dynamical structure of many complex systems from the internet
trough the biosystems to the social systems forms nontrivial
networks such as scale-free networks.  Therefore the dynamical
behaviors of the interacting (linked) units (nodes) on the
networks are very important to understand the dynamics of such
complex systems. Recently an important theory
\cite{MB1,MB2,EK1,EK2,EK3} on the dynamical behaviors on such
interacting (correlated) nodes is suggested and confirmed on the
internet, World Wide Web(WWW), river networks and etc.

The main theory \cite{MB1,MB2} for the dynamics of the node is
that there exists the power-law relation between the average
activity or flux $\left<f_i\right>$ of each node $i$ to the flux
fluctuation $\sigma_i$ as
\begin{equation}
\label{eq1} \sigma_i \sim \langle f_i \rangle ^\alpha.
\end{equation}
Furthermore the theory suggested that there can be two distinct
classes of dynamical systems. One class consists of the systems
with $\alpha \simeq 1/2$ and the other consists of those with
$\alpha=1$. The $\alpha \simeq 1/2$ systems are those in which the
main controlling dynamics is the internal intrinsic dynamics.
Examples are signal activity in microprocessors and time-resolved
information in internet routers. The $\alpha=1$ systems are those
in which the main dynamics is controlled by the external
variations. Examples are river networks, highway traffic and World
Wide Webs. Recently another study \cite{EK1} suggested a dynamical
model which has nonuniversal $\alpha$ values on complex networks
by introducing an impact variable to each node.

 To understand the origin of two distinct classes, several
dynamical models have been suggested \cite{MB1,MB2}. One of them
is based on a multi-random-walker model. The details of the
multi-random-walker model are as follows. In each experiment or
trial $d$, which was called as "day" in Refs. \cite{MB1,MB2}, the
number of unbiased random walkers $N_{RW}$ are randomly chosen
from the uniform distribution $[\langle N_{RW} \rangle -\triangle
N_{RW}, \langle N_{RW} \rangle +\triangle N_{RW} ]$. The walkers
are initially randomly distributed on a given support (network)
with the $N$ nodes (sites). Then the distributed walkers
simultaneously take preassigned fixed $T_{RW}$ steps in a given
experiment $d$ and count the number of visits (or flux) $f_{id}$
to the node $i$ during $T_{RW}$ steps. By repeating these
experiments $D$ times, one can get a series of data $\{f_{id}\}$
with $d=1,2,...,D$ and $i=1,2,...,N$. From the data we can
calculate the average flux and the flux fluctuation of a node $i$
as
\begin{eqnarray}
 \langle f_i (T_{RW}) \rangle = \frac{1}{D} \sum_{d=1}^D f_{id}
(T_{RW})~~,\\
 \sigma_i^2 (T_{RW}) = \frac{1}{D} \sum_{d=1}^D f_{id}^2
(T_{RW})- \langle f_i(T_{RW})\rangle ^2. \label{sig}
\end{eqnarray}
>From this model on the scale-free networks with the degree
exponent $\gamma=3$ the power-law relation $ \sigma_i (T_{RW})
\simeq \langle f_i (T_{RW})\rangle ^\alpha$ or Eq. (1) was
numerically shown. Furthermore the systems for small $\Delta
N_{RW}$ were shown to belong to $\alpha \simeq 1/2$ class, while
those for large $\Delta N_{RW}$ were shown to satisfy $\alpha=1$.
>From these results it was explained why $\alpha=1$ systems are
governed by the external flux variations and the behavior for
 $\alpha=1/2$ is from the internal intrinsic dynamics
 \cite{MB1,MB2}.

   In the multi-random-walk model, the flux configuration
$\{f_{id} (T_{RW})\}$ exactly corresponds to the height
configuration $\{h_{id}(T_{RW})\}$ in the following surface growth
model. In the growth model each of $N_{RW}$ random walkers drops a
particle on the node $i$ whenever it visits the node $i$ and thus
$f_{id}(T_{RW}) = h_{id} (T_{RW})$. Recently this kind of surface
growth model in which $N_{RW}$ is fixed (not-varied) in any
experiment $d$ is studied to know the effect of random-walk-like
colored noises \cite{SK} on the dynamical scaling properties of
the surface roughening \cite{dyn}. In the study \cite{SK} the
support is an ordinary one-dimensional lattice and the dynamical
scaling of the surface width $W$ satisfies the conventional
scaling as \cite{dyn}
\begin{equation}
W= L^{\alpha_s} f(T_{RW}/L^z) =  \begin{cases} T_{RW}^\beta, &
T_{RW} \ll L^z \\
L^{\alpha_s} , & T_{RW} \gg L^z
\end{cases}
\end{equation}
 with $z \equiv \alpha_s /\beta$.

In this paper we want to introduce another method to discriminate
the systems governed by the internal dynamics from those derived
by external input flux variations by comparison to corresponding
surface growth dynamics. The main quantity for the comparison is
the averaged $\sigma_i^2 (T_{RW})$ over the node $i$, which is
defined as
\begin{equation}
{\overline{\sigma}^2} (T_{RW}) \equiv \frac{1}{N} \sum_{i=1}^N
\sigma_i^2 (T_{RW}).
\end{equation}
As we shall see, $\overline{\sigma}(T_{RW})$ for the
multi-random-walker model shows the power-law dependence on the
walk step $T_{RW}$ as $\overline{\sigma}(T_{RW})\simeq
T_{RW}^\beta$. Furthermore, if the governing dynamics is not
anomalous but the internal intrinsic dynamics, then
$\overline{\sigma} (T_{RW})$ essentially has the same dynamical
behavior as that of the mean-square surface fluctuation (width)
$W(T_{RW})$ in one-random-walker surface growth model. Then
$T_{RW}$ is (acts as) the growth time and the exponent $\beta$ is
(or act as) the growth exponent in dynamical surface scaling
\cite{dyn}. If $W(T_{RW})$ in one-random-walker model dynamically
behaves as $W(T_{RW}) \simeq T_{RW}^{\beta_i}$, then
$\overline{\sigma} (T_{RW})$ for the systems governed by the
internal intrinsic dynamics satisfies $\overline{\sigma} (T_{RW})
\simeq T_{RW}^\beta$ with $\beta = \beta_i$. If $\beta \neq
\beta_i$, then governing dynamics should come from the external
flux variations.

We first want to show the exact mathematical relation of
$\overline {\sigma}$ to the surface fluctuation $W$. From Eq.
(\ref{sig}),
\begin{eqnarray}
\overline{\sigma}^2 (T_{RW}) &=& \frac{1}{N} \sum_{i=1}^N
\sigma_i^2 (T_{RW})
\nonumber\\
&=& \frac{1}{N} \sum_i \left [ \frac{1}{D} \sum_{d=1}^{D} f_{id}^2
 - \left( \frac{1}{D} \sum_{d=1}^{D}  f_{id}\right)^2  \right].
\end{eqnarray}
Since $f_{id} = h_{id}$ in the corresponding surface growth model,
the mean-square surface width $W^2$ is
\begin{eqnarray}
W^2(T_{RW}) &=& \frac{1}{D} \sum_{d=1}^D W_d^2 (T_{RW}) \nonumber \\
&=& \frac{1}{D} \sum_{d} \left [ \frac{1}{N} \sum_{i=1}^{N}
f_{id}^2 - \left( \frac{1}{N} \sum_{i=1}^{N} f_{id} \right)^2
\right] .
\end{eqnarray}
Thus
\begin{eqnarray}
\overline{\sigma}^2 (T_{RW})-W^2(T_{RW}) = \frac{1}{D} \sum_d
\left(
\frac{1}{N} \sum_i f_{id}\right)^2 \nonumber \\
-\frac{1}{N} \sum_i \left( \frac{1}{D} \sum_d f_{id}\right)^2
\label{dif}.
\end{eqnarray}
The difference between $\overline{\sigma}^2$ and $W^2$ for the
given step $T_{RW}$ comes from the different orders of taking the
average over nodes (i.e., $1/N \sum_i f_{id}(T_{RW}))$ and that
over experiments (or trials) (i.e., $1/D \sum_d f_{id} (T_{RW})$).

In mormal processes where any particular node does not receive
extraordinarily smaller or larger flux than any other regions,
then one can expect that the order of taking averages makes no
anomalous effects and thus $1/N \sum_i f_{id}(T_{RW}) \simeq 1/D
\sum_d f_{id} (T_{RW})$. Then the difference $\overline{\sigma}^2
(T_{RW})-W^2(T_{RW})$ becomes negligible or $\overline{\sigma}^2
(T_{RW})\simeq W^2(T_{RW})$. Therefore the possible origins which
make the physically important difference comes from the breakdown
of the spatial or temporal symmetry of the given system. Typical
examples for such breakdowns are the local columnar defect
\cite{dyn,Hal} and the temporally colored noises \cite{dyn,Med}.
When the input flux fluctuation becomes large or $\Delta N_{RW}$
is large in the multi-random-walker model, it is also possible
that the input flux to some localized region of the support can
become anomalously larger or smaller and then the difference
$\overline{\sigma}^2 (T_{RW})-W^2(T_{RW})$ can have some crucial
effects and remains for a certain length time.

To see the explained effects numerically, we analyze the numerical
data from the simulations for the random-walker models. To see the
universal features on homogeneous networks as well as
inhomogeneous networks, we use regular lattices of one and two
dimensions as well as random network(RN) \cite{AB} and scale-free
network (SFN) with $\gamma=3$ \cite{AB} as supports in the
simulations. Used size or the number $N$ of nodes (sites) of RN,
SFN, and an one-dimensional lattice are $10^4$ ($N=10^4$). The
used size of the two-dimensional square lattice is $100 \times
100$. Each experiment (trial) is repeated over $D=100$ times.
Every walker in each experiment $d$ takes $10^4$ steps. Data for
the flux $\{f_{id}(T_{RW})\}$ are obtained for the walk steps
$T_{RW}=1,2,...,10^4$.

The first simulation data to analyze are those for
one-random-walker model in which only one random walker is
initially dropped on a randomly chosen node (site) of the support
in a given experiment. In this model the flux dynamics should be
governed by the internal intrinsic dynamics and the difference
$\overline{\sigma}^2 (T_{RW})-W^2(T_{RW})$ is expected to be
negligible as explained in the previous paragraph. In Fig. 1 we
have plotted the surface fluctuation $W(T_{RW})$ and the average
flux variation $\overline{\sigma}(T_{RW})$ in one-random-walker
model. On the four kinds of support (RN, SFN, one and two
dimensional lattices) the dependence of
$\overline{\sigma}(T_{RW})$ on $T_{RW}$ (triangles in upper plots
of Figs. 1(a)-(d)) is almost the same as that of $W(T_{RW})$
(straight lines in the same corresponding figures.). We can't find
any significant difference $\overline{\sigma}^2
(T_{RW})-W^2(T_{RW})$. Furthermore the data for one random walker
model satisfies the scaling ansatz $\overline{\sigma}
(T_{RW})=W(T_{RW}) = T_{RW}^{\beta_i}$ very well  on both
inhomogeneous networks (RN and SFN) and homogeneous networks
(regular lattices). The estimated growth exponents $\beta_i$ for
one-random-walker model are $\beta_i=0.53(2)$ for SFN,
$\beta_i=0.51(1)$ for RN, $\beta_i=0.763(1)$ for one dimensional
lattices and $\beta_i=0.56(1)$ for two dimensional lattices. On
the inhomogeneous networks such as SFN and RN, $\beta_i$ is very
close to $\beta=1/2$, which is the value of $\beta$ of random
ballistic depositions on any support \cite{dyn}. In contrast on
one and two dimensional lattices $\beta_i$ is quite different from
1/2.

\begin{figure}
\includegraphics[scale=0.5]{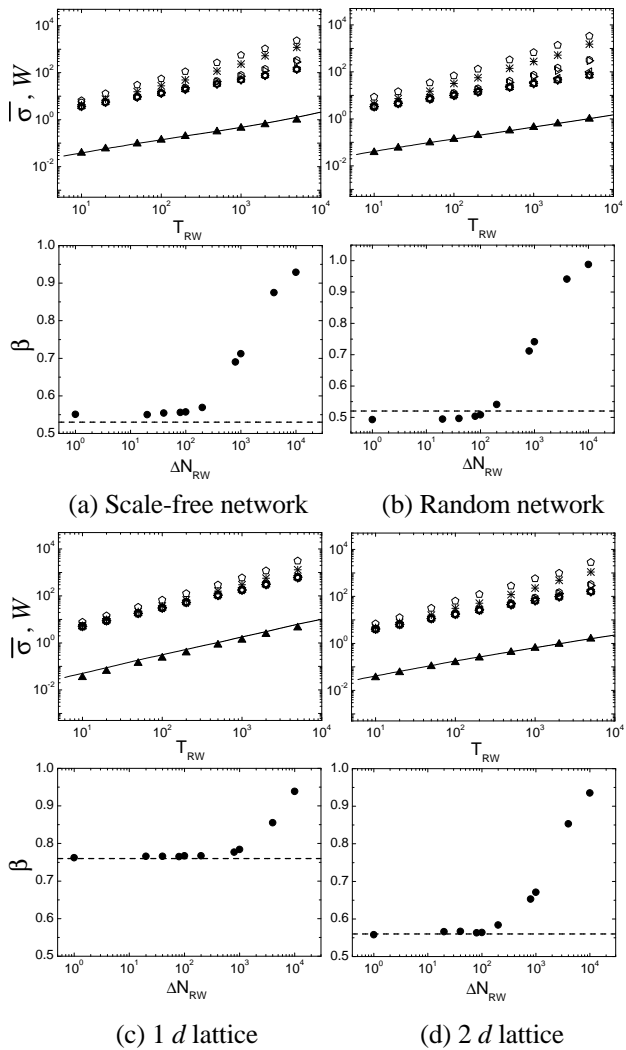}
\caption{ Triangles ($\blacktriangle$) and straight lines in the
upper plots of figures (a), (b), (c), and (d) represent the data
for the average flux fluctuation $\overline{\sigma}(T_{RW})$ over
nodes and the surface fluctuation $W(T_{RW})$ in the
one-random-walker model, respectively. Other symbols in the upper
plots represent $\overline{\sigma}(T_{RW})$ for $\Delta N_{RW}=0,
20, 40, 80, 100, 200, 800, 1000, 4000, 10000$ from bottom to top
in the multi-random walker model. The lower plots show the values
of exponent $\beta$ for various $\Delta N_{RW}$ obtained by
fitting the data in upper plots to the relation
$\overline{\sigma}(T_{RW}) \simeq T_{RW}^\beta$. The dotted lines
in the lower plots denote the values of $\beta_i$ obtained from
the relation $W(T_{RW}) \simeq \overline{\sigma} (T_{RW}) \sim
T_{RW}^{\beta_i}$ in one-random-walker model. (a) on Scale-free
network with $\gamma=3$. (b) on Random network. (c) on one
dimensional lattice. (d) on two dimensional lattice.}
\end{figure}

In the simulations for multi-random-walker models, the number
$N_{RW}$ of walkers in an experiment $d$ is selected randomly
among the uniform distribution $\left[\left<N_{RW}\right> - \Delta
N_{RW},\left<N_{RW}\right> +\Delta N_{RW}\right]$. Then the
$N_{RW}$ walkers are initially randomly distributed over the nodes
(sites) of the support. In the simulations $\left<N_{RW}\right>=
10^4$ is always imposed and $ \Delta N_{RW}$ is varied as $\Delta
N_{RW} = 0, 20,40, 80, 100, 200, 800, 10^3, 4 \times 10^3, 10^4$.
Other conditions are exactly the same as those in the simulations
for one-random-walker model. The results for the dependence of
$\overline{\sigma}$ on $T_{RW}$ in multi-random-walker model for
various $ \Delta N_{RW}$ are also shown in upper plots of Figs.
1(a)-(d). From the data and the relation $\overline{\sigma} \simeq
T_{RW}^\beta$ the obtained values of $\beta$ for $\Delta N_{RW}$'s
are displayed in the lower plots of Figs. 1(a)-(d), where the
dotted line denotes the values of $\beta_i$ in one-random-walker
model on the corresponding support. As can be seen from the lower
plots, $\beta$ value for small $\Delta N_{RW}$ or $\Delta N_{RW}
\le 200$ is nearly equal to $\beta_i$ (i.e., $\beta=\beta_i$) on
both homogenous and inhomogeneous networks. On the other hand
$\beta$ for large $\Delta N_{RW}$ or $\Delta N_{RW} \ge 4 \times
10^3$ approaches to $\beta =1$ (or $\beta \ge 0.9$). $\beta$ for
crossover values of $\Delta N_{RW}$ (or $\Delta N_{RW} = 800$ or
$1000$) has crossover values between $\beta_i$ and 1. The results
for the multi-random walker models are as follows. For small
$\Delta N_{RW}$ or in the systems where the internal intrinsic
dynamics is dominant, $\overline{\sigma} (T_{RW})$ follows the
behavior of $W (T_{RW}) \simeq T_{RW}^{\beta_i}$ in
one-random-walker model quite well. For large $\Delta N_{RW}$ or
in the systems where the external flux fluctuation is dominant for
dynamical behaviors, $\overline{\sigma} (T_{RW})$ follows the
behavior $\overline{\sigma} (T_{RW}) \simeq T_{RW}^\beta$ with
$\beta=1$.

In summary, we have shown that $\overline{\sigma}$ shows the
power-law dependence on the walk step $T_{RW}$ as
$\overline{\sigma} \simeq T_{RW}^\beta$ in multi-random-walker
model for various $\Delta N_{RW}$. To discriminate the systems
governed by the internal intrinsic dynamics from those governed by
the external flux variations, the dynamical behavior of
one-random-walker model is suggested as the criterion. If the
dependence of $\overline{\sigma}$ in multi-random-walker model on
$T_{RW}$ follows the dynamical behavior of the surface width in
one-random-walker model as $W(T_{RW}) \simeq T_{RW}^{\beta_i}$ or
$\beta=\beta_i$, then the model is governed by the internal
dynamics. This behavior occurs when $\Delta N_{RW}$ is small. When
$\beta \neq \beta_i$ and $\beta=1$, then the model is governed by
the external flux variations. The $\beta=1$ behavior occurs when
$\Delta N_{RW}$ is large. We also show that this criterion holds
for the model not only on the inhomogeneous networks but also on
homogeneous networks or regular lattices. Furthermore $\beta_i$
for one and two dimensional lattices is greater than 1/2.
Especially in one dimension the random walker has the temporal
correlation due to the reentrant property and the visitation
probability to a certain site has some power-law correlation in
time \cite{SK,Red}. In two-dimension the reentrant property is
marginal \cite{Red}. This reentrant property of random walker
explains $\beta_i > 1/2$ in both one and two dimensional lattices.
In contrast for SFN and RN whose effective dimensionality is
$\infty$ \cite{AB}, the random walkers are transparent \cite{Red}
and the one-random-walker model behave as random ballistic
deposition and satisfies $\beta=1/2$ \cite{dyn}.

\begin{figure}
\includegraphics[scale=0.5]{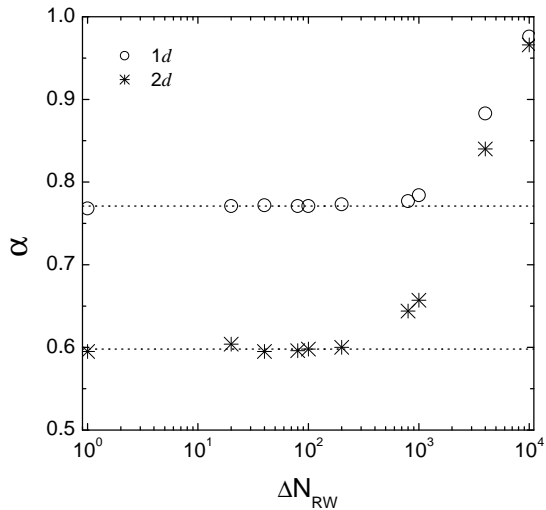}
\caption{Dependence of the exponent $\alpha$ on $\Delta N_{RW}$ of
multi-random-walker model in one ($1d$) and two ($2d$) dimensional
lattices. The data is obtained by fitting $ \sigma_i (T_{RW})$
 versus $\left<f_i ({T_{RW}})\right>$  to the relation (1) or $
\sigma_i (T_{RW}) \simeq \left<f_i (T_{RW})\right>^\alpha$. The
data for $ \sigma_i (T_{RW})$ and $\left<f_i(T_{RW})\right>$ in
one and two dimensions are obtained from the same simulation
conditions as those in Figs. 1(c) and (d) with a fixed value of
$T_{RW}$ as $T_{RW}=100$.}
\end{figure}

Final comments are on the nonuniversal values of exponent $\alpha$
for Eq. (1) in regular lattices or homogeneous networks. Recently
a study \cite{EK1} suggested a model in which nonuniversal
$\alpha$ values (i.e., $\alpha \neq 1 ~ \rm{or} ~ 1/2$) for the
relation $ \sigma_i (T_{RW}) \simeq \left<f_i
(T_{RW})\right>^\alpha $ with the fixed $T_{RW}$. By using the
multi-random-walker model with the quenched impact variable $V(i)$
for each  node $i$, which is dependent on the degree $k(i)$, the
study \cite{EK1} showed nonuniversal $\alpha$ values (i.e. $ 1/2 <
\alpha < 1$) on scale-free networks. Nonuniversal $\alpha$ values
are also possible on the homogeneous networks or regular lattices
without the impact variable $V(i)$. By using the relation $
\sigma_i (T_{RW}) \simeq \left<f_i (T_{RW})\right>^\alpha $ with
the fixed $T_{RW}$ on one and two dimensional lattices, we obtain
$\alpha$ values for various $\Delta N_{RW}$. The results are shown
in Fig. 2. The simulation conditions except $T_{RW}$ for Fig. 2
are exactly the same as those for Figs. 1(c) and (d). $T_{RW}$ for
each experiment (day) is fixed as $T_{RW}=100$. For $\Delta N_{RW}
< 500$ where the internal intrinsic dynamics are dominant,
$\alpha$ values are nearly constant. The values of the exponent
$\alpha$ for small $\Delta N_{RW}$ are $\alpha=0.77(1)$ in one
dimension and $\alpha = 0.60(2)$ in two dimension, respectively.
The behavior $\alpha > 1/2$ for internal intrinsic dynamics should
come from the reentrant properties of random walks. As explained
before the reentrant property makes the power-law correlation in
time for the probability to visit a certain site again. This kind
of temporal correlation effects makes the probability distribution
for $\left<f_i\right>$ deviate from the Gaussian and thus gives
nonuniversal $\alpha$ values for small $\Delta N_{RW}$. For large
$\Delta N_{RW}$ $\alpha$ approaches 1 even for homogeneous
networks as can be seen from data for $\Delta N_{RW} = 4000,
10000$ in Fig. 2.

This work was supported by Grant No. R01-2004-000-10148-0 from the
Basic Research Program of KOSEF. We thank Dr. Soon-Hyung Yook  for
valuable suggestions. One of authors (YK) also acknowledges the
hospitality of Korea Institute of Advanced Study during the visit.


\begin{references}
\bibitem{AB} R.Albert and A.-L.Barab\'{a}si, Rev.Mod.Phys. {\bf 74},
47 (2002); S.N.Dorogovtsev and J.F.F.Mendes, Adv.Phys. {\bf 51},
1079 (2002).
\bibitem{MB1} M. Argollo de Menezes and A. -L. Barab\'asi,
Phys. Rev. Lett. {\bf 92}, 028701 (2004).
\bibitem{MB2} M. Argollo de Menezes and A. -L. Barab\'asi,
Phys. Rev. Lett. {\bf 93}, 068701 (2004).
\bibitem{EK1} Z. Eisler and J. Kert\'ezes, Phys. Rev. E {\bf 71}, 057104 (2005).
\bibitem{EK2}  Z. Eisler and J. Kert\'ezes, S. -H. Yook, and
A. -L. Barab\'asi, arXiv:cond-mat/0408409 (2004).
\bibitem{EK3}  Z. Eisler and J. Kert\'ezes, arXiv:cond-mat/0503139 (2005).
\bibitem{SK} H. S. Song and J. M. Kim, Jour. of the Korean Phys.
Soc. {\bf 44}, No. 3, 543 (2004).
\bibitem{dyn}
       {\em Dynamics of Fractal Surfaces},
        edited by F. Family  and  T. Vicsek
        (World Scientific, Singapore, 1991);
     A.-L. Barab\'{a}si and H. E. Stanley,
      {\em Fractal Concepts in Surface Growth} (Cambridge University
       Press, Cambridge, 1995); J. Krug, Adv. Phys. {\bf 46}, 139 (1997).
\bibitem{Red}
       S. Redner, {\em A GUIDE TO FIRST PASSAGE PROCESSES}
     (Cambridge University Press, Cambridge, 2001).
\bibitem{Hal} T. Halpin-Healy and Y.-C Zhang, Phys. Rep. {\bf
254}, 215 (1995).
\bibitem{Med} E. Medina, T. Hwa, M. Kardar and Y.-C Zhang,
Phys. Rev. A {\bf 39}, 3053 (1989).




\end{references}
\end{document}